\documentclass{article}  
\usepackage{spadre2008}
\usepackage{graphicx}
\frompage{000} \topage{000}                                              %|
\def\rightmark{A Personal Account of Some PHOBOS Physics}
\def\leftmark{W. Busza}
 
\title{A Personal Account of Some PHOBOS Physics} 
\authors{
{Wit Busza
}\\[2.812mm]
{\normalsize
\hspace*{-8pt}Massachusetts Institute of Technology 
Cambridge MA USA\\[0.2ex] 
}}
 
\abstract{This paper discusses some aspects of PHOBOS physics and its origins, in particular participant scaling and extended longitudinal scaling, seen in A+A and h+A collisions at all energies.}

 
\begin{document}
 
\maketitle
\setcounter{page}{1}

With the PHOBOS project {\cite{bib1}} winding down it seems a good time to discuss some of the more intriguing PHOBOS results and their origins, in particular $N_{part}$ scaling and extended longitudinal scaling (a.k.a. limiting fragmentation) in h+A and A+A collisions.  The fact that the number of participants is a relevant parameter in describing hadronic collisions and that rapidity or pseudorapidity distributions exhibit extended longitudinal scaling, are topics that have fascinated me for more than 30 years.  Today participant scaling and extended longitudinal scaling, even for the elliptic flow parameter $v_2$, are so commonly known that they are taken for granted, and very few people stop and ask ``why''.  Yet it is important to remember that this simplicity and universality exhibited by data is far from obvious and far from being understood.

In the late 1960s and early 1970s there was much interest in the mechanism of multiparticle production{\cite{bib2}}.  Some people, including myself, were particularly interested in the space-time evolution of the production process{\cite{bib3}}.  An obvious way to probe the evolution process was to study how the number of produced particles depends on the nuclear thickness in h+A collisions.  Crudely, if the evolution was a slow process it would lead to little multiplication of particles in the nucleus.  If it was a fast process, it would lead to an exponential growth with thickness.  The parameter $\bar{\nu}$  was introduced, which measures the average thickness in units of the mean free path of hadrons (conventionally the incident hadron) in nuclear matter.  It was a very convenient parameter since it can be measured:   $\bar{\nu} = \frac{\sigma_{hp}A}{\sigma_{hA}}$ where $\sigma_{hp}$ and $\sigma_{hA}$ are the h+p and h+A inelastic cross-sections.  In today's language, of course,  $\nu = N_{part}-1$.

At first, multiparticle production in h+A collisions was studied using the nuclear emulsion technique{\cite{bib4}}.  A difficulty of interpretation of emulsion data is the fact that in any one collision it is not possible to unambiguously identify the target nucleus.  In the late 60s and early 70s an ingenious cosmic ray experiment was carried out at Echo Lake by Larry Jones and his collaborators{\cite{bib5}}.  Observing cosmic ray collisions in Al, Fe, Sn, and Pb plates in the energy range $\sqrt{s_{NN}} = 13-31 GeV$, they concluded that the multiplicity of produced particles increased very slowly with A; the total multiplicity increasing approximately logarithmically with A and the very forward production almost not at all.  These results encouraged me to carry out a systematic study of the $\nu$ or A-dependence of particle production at high energies using the newly constructed accelerator at Fermilab.  It led to the E178 experiment {\cite{bib6}}(which I like to call ``PHOBOS 1") and its upgrades.  In E178{\cite{bib7}} for 50-200 GeV $\pi$, k, p beams we observed for the first time that the cascading of particles inside nuclei is best described by $R=\frac{n_{hA}}{n_{hp}}=\frac{1}{2}+\frac{1}{2}\bar{\nu} = \frac{N_{part}}{2}$, that 
later became known as participant scaling.  It is important to point out that this was contrary to nearly everyone's expectations{\cite{bib3}}.  At the time the conventional wisdom was that long range correlations will not be important and so as $E \rightarrow \infty$, $R \rightarrow 1$.  One of few exceptions was Gottfried's energy-flux cascade{\cite{bib8}}, a model with some resemblance to current thinking, which predicted that at high energies $R= 0.62 + 0.38 \bar{\nu}$.
I repeat, the observed $N_{part}$ scaling was anything but obvious to those in the field.  It led Bia{\l}as et al {\cite{bib9}} to introduce the phenomenological idea of a ``wounded nucleon" and to the wounded nucleon model.  

h+A studies in E178 and also p-Emulsion studies made other observations that attracted attention.  One was that limiting fragmentation of the target (energy independence and projectile independence) also applied to h-A collisions and not just to h-p collisions{\cite{bib10}}.  This was not too surprising.  It basically followed from the prevailing parton picture of the time.  It was beautifully explained by A. Goldhaber{\cite{bib11}}.  He considered the production of particles in different frames of reference and took into account the formation time of particles, to qualitatively explain the facts.  

Another observation {\cite{bib10}} and much more surprising, was that the region of rapidity which exhibited limiting fragmentation increases with energy, a phenomenon not well understood to the present day and which recently Mark Baker named ``extended longitudinal scaling".  

A third related phenomenon, and equally not well understood, was the fact that wherever extended longitudinal scaling holds, the ratios of produced particles are independent of the target nucleus{\cite{bib12}}.  

The advent of data from relativistic heavy ion collisions added a new dimension to the study of particle production.  Here the hadronic collisions occur in close proximity to each other in a large volume.  The probability of reinteraction of the initially produced particles or matter is high.  If reinteractions do occur and if a new intermediate state is created, naively one would expect that the simple scalings observed in p+A, $N_{part}$ scaling and extended longitudinal scaling, should breakdown.  such a breakdown would in fact be a signature of the onset of new physics and thus extremely interesting.  Naively one would also expect that in A+A, as a result of reinteractions, there would be significant increase in the total multiplicity.  Furthermore, as a consequence of the particle increase and the fact that the overall volume of the produced interacting system is so large it would be reasonable to expect a decrease in the transverse momenta of the particles.  

It is such considerations that led to the design of the PHOBOS multiplicity array{\cite{bib1}}, and it is no accident that it resembles so closely the E178 detector{\cite{bib6}}.  See Figure 1.

\begin{figure}[htbp] %  figure placement: here, top, bottom, or page
   \centering
   \includegraphics[width=4in]{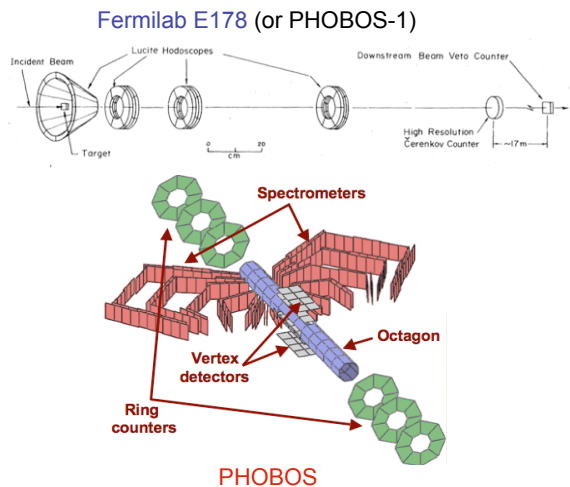} 
  \caption{Top: Detector of Experiment E178 at Fermilab which in the 1970s studied particle production in $\pi$, k, p+A collisions.  Bottom: The PHOBOS detector at RHIC.}
   \label{fig1}
\end{figure}

The other main component of the PHOBOS detector, the two arm multiparticle spectrometer, reflects the influence of the results obtained with heavy ion collisions at low energies, at the AGS and SPS.  Physics obtained with this part of PHOBOS is another story and will not be discussed here.

The strongly interacting system created at RHIC was the major surprise for the community.  For me a no lesser surprise was how similar all the global trends in A+A were to those seen in h+A collisions.  PHOBOS spent a large fraction of its effort studying the global features and it is good to reflect on these results.  

There is no significant increase in particle production at RHIC.  The mid-rapidity particle density increases smoothly and only logarithmically with $\sqrt{s_{NN}}${\cite{bib13}}.  There is no increase of the production of low $p_t$ particles{\cite{bib14}}.  The centrality dependence is weak and independent of energy.  The baryon rapidity loss is the same as in pA collisions{\cite{bib15}}.  The production mechanism seems to be governed more by the geometry of the collision than by the system size.  Above all, $N_{part}$ scaling seems to hold in A+A collisions as it held in h+A collisions{\cite{bib16}}.  So does extended longitudinal scaling{\cite{bib17}}, which even seems to apply to the elliptic flow!{\cite{bib18}}

It is possible that this simplicity in the global data seen from pp through pA to AA collisions, from $\sqrt{s_{NN}}=10 GeV$ to $\sqrt{s_{NN}} = 200 GeV$, has some trivial origin.  For example it might follow from the fact that in the phase diagram there is a cross-over or because the results are dominated by  saturation effects in the early stages of the collision.  However it is also possible that these results are trying to tell us something profound, which so far we have failed to understand.  We must not overlook this possibility.  To help us understand these phenomena we may soon get another clue when LHC turns on.

It will be fascinating to see if the data at LHC exhibit the simplicity seen at lower energies, and if trends seen at lower energies continue to persist to these extremely high energies.  Current theoretical prejudice{\cite{bib19}} is that multiplicities will be higher and flow weaker than predictions based on simple extrapolation of RHIC data to LHC energies, using participant scaling, extended longitudinal scaling, and logarithmic rise of mid rapidity particle density{\cite{bib20}}.  

As a conclusion, it is a sobering thought that for more than 30 years I have tried to understand the origin of participant scaling and extended longitudinal scaling.  As we study the collisions of more and more complicated systems, at higher and higher energies, at every step it seems obvious that these scalings should breakdown, and yet they insist on persisting.  Today I feel no closer to my goal than I was at the beginning!

\vfill\eject
\end{document}